\documentclass[manuscript]{aastex}

\shorttitle{The Unique Eclipses of KH 15D}
\shortauthors{Herbst et al. }

\begin{document}


\title{Fine Structure in the Circumstellar Environment of a Young, 
Solar-like Star: the Unique Eclipses of KH 15D}


\author{William Herbst and Catrina M. Hamilton\altaffilmark{1}}
\affil{Astronomy Dept., Wesleyan U., Middletown, CT 06459}

\author{Frederick J. Vrba}
\affil{U.S. Naval Obs., Flagstaff Station, Box 1149, Flagstaff, 
AZ 86002-1149}

\author{Mansur A. Ibrahimov}
\affil{Ulugh Beg Astronomical Institute of the Uzbek Academy of 
Sciences, Astronomicheskaya 33, 700052 Tashkent, Uzbekistan}

\author{Coryn A. L. Bailer-Jones, Reinhard Mundt and Markus Lamm}
\affil{Max-Planck-Institut f\"ur Astronomie, K\"onigstuhl 17, 
D-69117 Heidelberg, Germany}

\author{Tsevi Mazeh}
\affil{School of Physics and Astronomy, Raymond and Beverly Sackler 
Faculty of Exact Sciences, Tel Aviv University, Tel Aviv, Israel}

\author{Zodiac T. Webster}
\affil{Astronomy Dept., U. C. Berkeley, 601 Campbell Hall, Berkeley, 
CA 94720-3411}

\author{Karl E. Haisch}
\affil{NASA Ames Research Center, Mail Stop 245-6, Moffett Field, 
CA 94035-1000}

\author{Eric C. Williams and Andrew H. Rhodes}
\affil{Astronomy Dept., Wesleyan U., Middletown, CT 06459}

\author{Thomas J. Balonek}
\affil{Dept. of Physics and Astronomy, Colgate U., 13 Oak Drive, 
Hamilton, NY 13346-1398}

\author{Alexander Scholz}
\affil{Th\"uringer Landessternwarte Tautenburg, Sternwarte 5, D-07778 
Tautenburg, Germany}

\and

\author{Arno Riffeser}
\affil{Universit\"ats-Sternwarte M\"unchen, Scheinerstr. 1,D-81679 
M\"unchen, Germany}


\altaffiltext{1}{Physics Dept., Connecticut College, Box 5424, New London,
CT 06320}


\begin{abstract}
Results of an international campaign to photometrically monitor the 
unique pre-main sequence eclipsing object KH 15D are reported. An updated 
ephemeris for the eclipse is derived that incorporates a slightly revised 
period of 48.36 d. There is some evidence that the orbital period is 
actually twice that value, with two eclipses occurring per cycle. The 
extraordinary depth ($\sim$3.5 mag) and duration ($\sim$18 days) of the 
eclipse indicate that it is caused by circumstellar matter, 
presumably the inner portion of a disk. The eclipse has continued to 
lengthen with time and the central brightness reversals are not as 
extreme as they once were. V-R and V-I 
colors indicate that the system is slightly bluer near minimum light. 
Ingress and egress are remarkably well modeled by the 
passage of a knife-edge across a limb-darkened star. Possible models 
for the system are briefly discussed.    
\end{abstract}




\section{Introduction}

Planets are believed to form during the first $\sim$100 My of a star's life 
from matter in a circumstellar disk \citep{cw98}. The initial stages of that 
process have been viewed with increasing clarity in high-resolution images of 
young stars obtained with the Hubble Space Telescope. These have revealed 
complex disk structures on scales of tens to hundreds of AU
around $\sim$1 My old objects \citep{o01,ksw}. Probing the inner parts 
of disks, where terrestrial planets formed in our solar system and where 
giant planets are found in many exo-solar systems \citep{mb00,v02},
however, is still well beyond the reach of the current generation of 
telescopes. Here we report observations of a remarkable eclipsing solar-like 
star at an age of $\sim$3 My, which is providing a glimpse of the structure 
and possibly evolution of circumstellar matter on scales as fine as 0.01 AU. 
This is possible because of its unique geometry that results in the star 
being periodically eclipsed by extended non-luminous matter in its vicinity. 
No other object in the history of astronomy has been found to behave in quite 
the same way. 

The star in question, KH 15D ($\alpha = 6^{h}41^{m}10.18^{s}$, 
$\delta = 9\degr 28\arcmin 35.5\arcsec$, epoch 2000; see \citet{h01} 
for an identification chart), was first noticed in 1997 as a unique and 
potentially important object during a photometric monitoring program of 
young clusters undertaken with a small telescope at Van Vleck Observatory 
(VVO) on the campus of Wesleyan University in Middletown, CT 
\citep{kehm,kh98}. It undergoes a very deep ($\sim$3.5 mag) eclipse every 48.3 days 
and remains in the faint state for about 18 days, suggesting that the 
eclipsing body is non-luminous circumstellar matter. Remarkably, there is a 
central reversal of variable height that sometimes returns the star to its 
out-of-eclipse level (and once to an even brighter level) for a brief time 
near mid-eclipse. A follow-up study by \citet{h01} revealed that the star has 
a mass of 0.5 - 1 solar masses, and a radius of about 1.3 solar radii, 
indicating that it is still in its contraction phase towards the main sequence 
and has not yet initiated hydrogen burning in its core. Its pre-main sequence 
status is confirmed by the presence of Li I in its spectrum and its membership 
in the young open cluster NGC 2264, which has an age of 2-4 My and a distance 
of 760 pc \citep{sbl}. No other star, among the thousands monitored in the 
VVO program or elsewhere over more than a decade \citep{smmv,hrhc,reb01,hbjm}
has behaved in similar fashion. The closest analogues we can find in the 
astronomical literature are $\epsilon$ Aur, an F-type supergiant which is 
eclipsed every 27 years by dark circumstellar matter \citep{l96}, and BM Ori, 
an early B star which is eclipsed every 6.5 days by something dark orbiting it
\citep{pp76}. Neither star is solar-like nor are the eclipses even remotely 
similar to those of KH 15D.

\section{Observations}

To investigate this phenomenon in more detail, we organized, during the 
2001/2002 observing season, an international campaign aimed at obtaining 
precision photometry with the highest possible time resolution over the 
longest possible time interval. A full report on this effort will be 
given elsewhere \citep{ham02}. Here we provide a brief overview of 
the phenomena and an ephemeris to facilitate observations. 
The principal sources of optical photometric 
data were telescopes of 0.6-1.5 m aperture at the U. S. Naval 
Observatory's (USNO) 
Flagstaff Observing Station in Arizona, Maidanak Observatory in Uzbekistan, 
Wise Observatory in Israel, Kitt Peak National Observatory (KPNO)
and VVO. Additional contributions, some at earlier 
epochs, came from Calar Alto Observatory in Spain, the European Southern 
Observatory in Chile and a campus telescope at Colgate University in New York. 
Differential photometry relative to a grid of local comparison stars was used 
to establish zero points and combine the data. Photometric accuracy 
is $\sim$0.02 mag out of eclipse but, because of the faintness of the star 
and its proximity to a much brighter cluster member, declines to $\sim$0.1 mag 
during eclipse. Most observations were obtained in the I band of the Cousins 
system but some other filters were used, as discussed below. We obtained some 
data on all five eclipses that occurred during the observing 
interval 11 Sept. 2001 to 3 April 2002; excellent coverage was obtained on 
the last three eclipses, in Dec., Jan.-Feb., and Mar., respectively.

\begin{figure}
\plotone{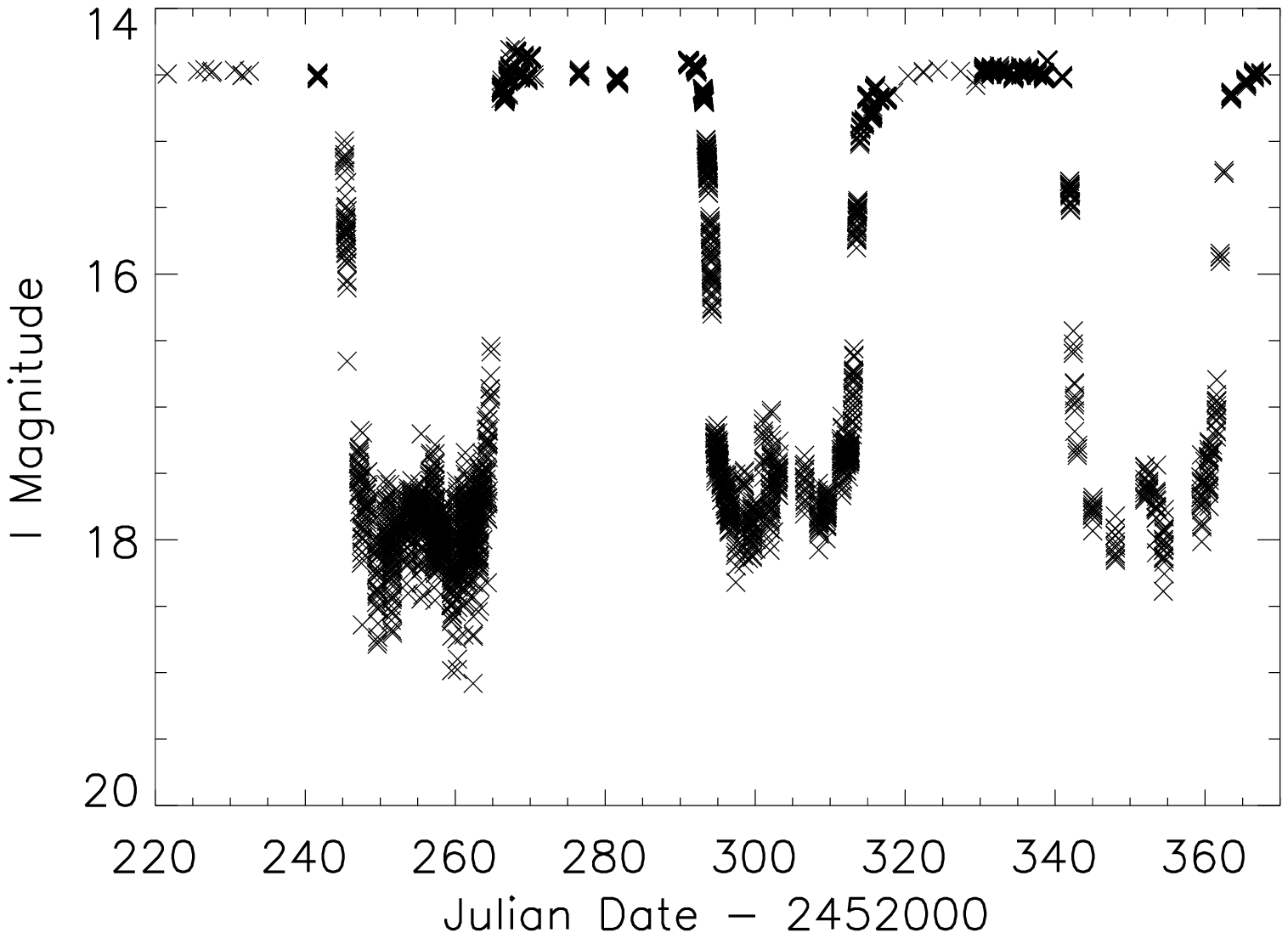}
\caption{A portion of the light curve of KH 15D during the 01/02  
observing season showing three eclipses. The last eclipse displayed 
above was centred at Julian Date 2452352.26. Based on these data we 
predict that future eclipses will occur at intervals of 48.36 days. 
The current width of the eclipses, measured at the 16.25 mag. level is 
almost exactly 0.4 in phase, or 19.3 days. Ingress and egress begin 
approximately 1.5 and 1.9 days before or after the 16.25 mag level, 
respectively. The rate of decline becomes much shallower about 0.5 days 
after that level is reached, and the deepest minimum may not occur until 
many days later. \label{lightcurve}}
\end{figure}

A portion of the light curve, including the three well-sampled eclipses of 
last season, is shown in Fig. \ref{lightcurve}. The basic features, including 
a rapid ingress and egress phase, a deep, structured minimum with central 
reversal, and an eclipse duration of about 40\% of each cycle are evident.
Combining these data with what 
has been observed over the six year interval since its discovery, we 
have derived a new ephemeris for mid-eclipse:

$$ \rm{JD (mideclipse)} = 2452352.26 + 48.36\rm{E}. $$  

The current width of the eclipses, measured at the 16.25 mag. level is 
almost exactly 0.4 in phase, or 19.3 days. Ingress and egress begin 
approximately 1.5 and 1.9 days before or after the 16.25 mag level, 
respectively. The rate of decline becomes much shallower about 0.5 days 
after that level is reached, and the deepest minimum may not occur until 
many days later. There were also subtle variations from eclipse 
to eclipse last season, as may be noticed by close inspection of Fig. 
\ref{lightcurve}. In fact, while this requires verification, 
it appears quite possible that 
the true period for this star is not the average interval between eclipses 
(48.36 days), but twice that, 96.72 days.  In particular, the ``central" 
reversals of alternate eclipses are slightly offset from mid-eclipse by 
similar amounts but in opposite direction while the adjacent minima are 
symmetric, but in mirror image fashion.

\begin{figure}
\plotone{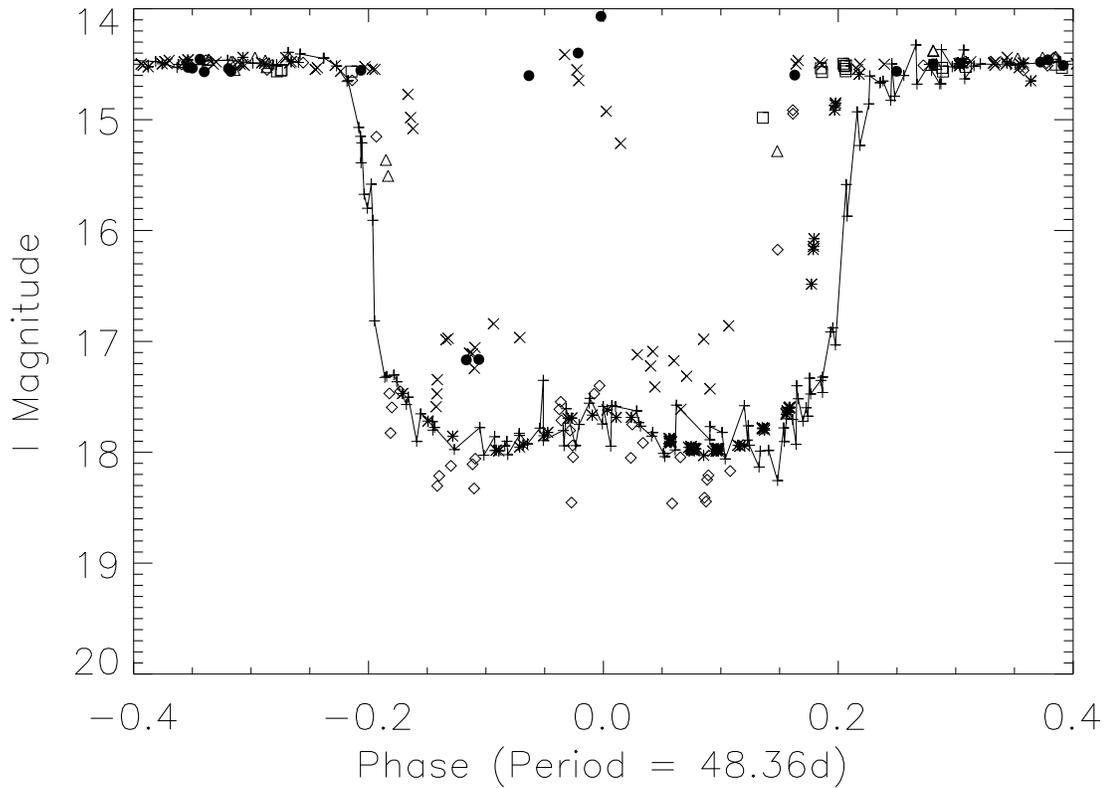}
\caption{The eclipses of KH 15D since its discovery in 1995, phased with 
the period 48.36 days. Different symbols refer to different years, as follows: 
95/96 (solid circle), 96/97 (x), 97/98 (square), 98/99 (triangle), 99/00 
(diamond), 
00/01 (asterisk), 01/02 (plus signs connected by lines). The data, 
averaged by day, 
from the 01/02 season and connected by the solid line clearly
define the outer limits of the observed points 
in phase. This shows that the eclipse has been widening with time. It 
is also noticeable that 
the central reversals have declined in brightness with time. \label{eclipse}}
\end{figure}

A second remarkable result from last season's intensive monitoring effort is 
that the breadth of the eclipse and height of the central reversal have 
continued to evolve in secular fashion (see Fig. \ref{eclipse}). Phasing all 
of the photometry obtained over the six year interval since its discovery with 
the 48.36 d period, we see that the most recent data define the outer limits 
of the eclipse in phase. That is, at earlier epochs, eclipses were clearly of 
shorter duration and the central reversals often reached much brighter levels, 
sometimes climbing to the out-of-eclipse level or, in one case, slightly 
brighter. Fig. \ref{eclipse} may represent the first detection of changing 
structure in the circumstellar disk of a young star and provides a serious 
challenge to models of the system. It is currently impossible to predict how 
the eclipses will evolve with time, but the fact that they are changing so 
dramatically on human time scales underscores the necessity to keep 
monitoring this star. 

\begin{figure}
\plotone{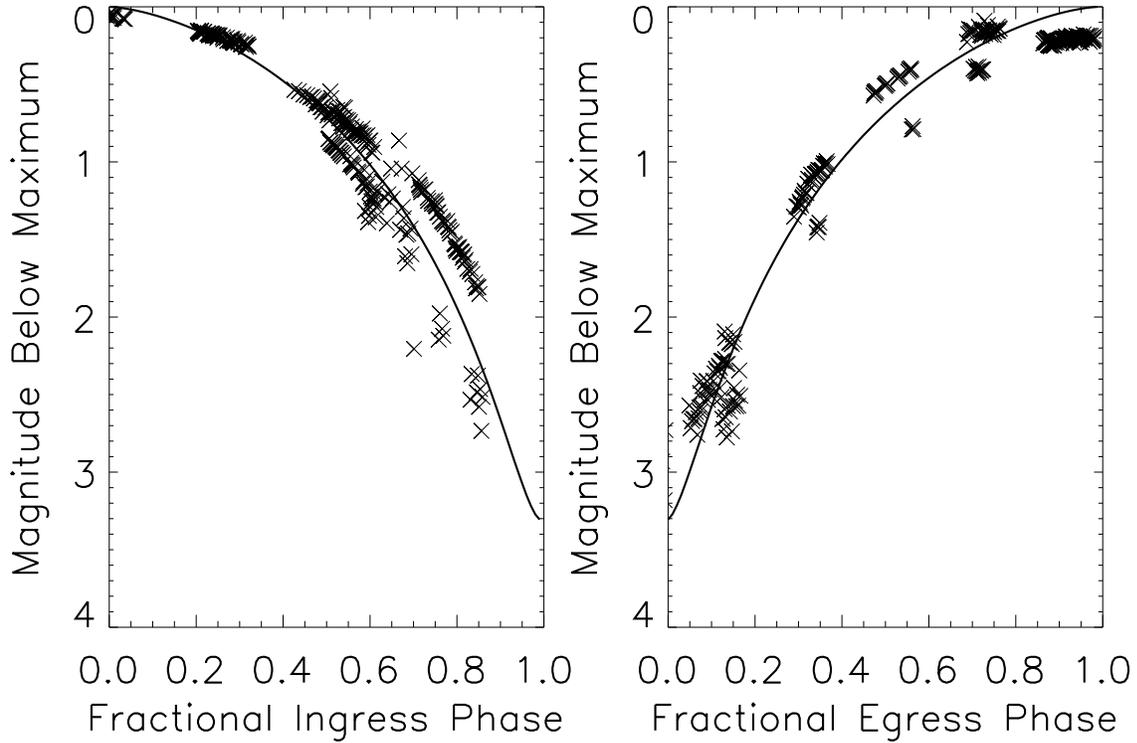}
\caption{Ingress and egress of all eclipses observed in 01/02 compared 
with a knife edge model. The model assumes a perfectly straight, sharp edge 
cutting across a limb-darkened star over a 1.9 day period for ingress and 2.4 
day period for egress. It includes a 5\% addition for scattered light in the 
system, which becomes important only near minimum light. Clearly the models 
are a reasonable representation of the data.\label{knifeedge}}
\end{figure}

Ingress and egress are remarkably smooth and have shapes that are well 
represented by the steady advance or retreat of a ``knife-edge" across a 
limb-darkened star. In the case of ingress this takes $\sim$1.9 d, while 
for egress it is slightly longer -- $\sim$2.4 d. A comparison between the 
knife-edge models and the data, phased with the 48.36 d period, is shown in 
Fig. \ref{knifeedge}. A standard limb-darkening law, 
$$I(\theta) = I(0) - \mu + \mu cos\theta$$ with $\mu=0.3$, appropriate to a 
star of effective temperature 4000 K, was assumed \citep{d87}. The shape of 
the model light curve is quite insensitive to the value of $\mu$. It derives 
primarily from the geometric feature that near the halfway point the edge is 
cutting longer chords of the star. It is important to note that the duration 
of ingress and egress are much longer than the expected transit time 
($\sim$0.5 d) of an object on a Keplerian orbit about the star with a period 
of 48 or 97 days. 

\begin{figure}
\plotone{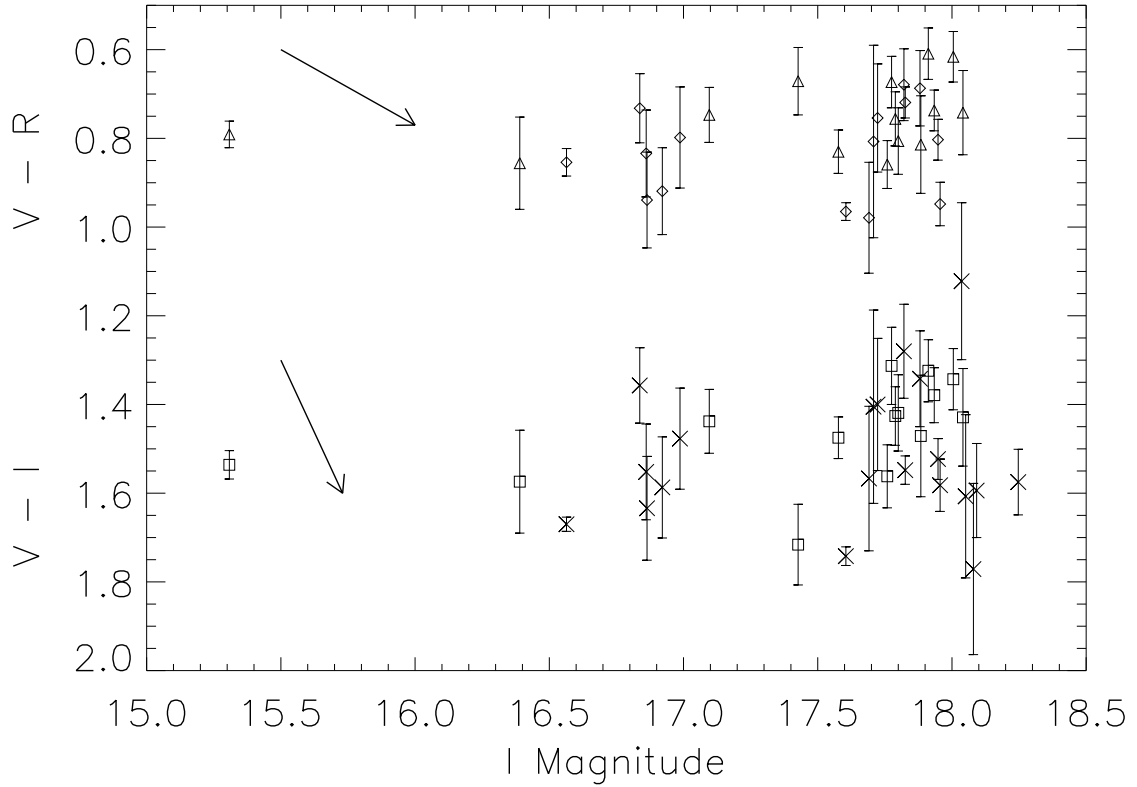}
\caption{Color versus brightness. In V-R, triangles indicate data 
obtained at the Flagstaff station of the USNO and diamonds indicate data 
obtained at KPNO. In V-I, USNO data is indicated by squares and KPNO 
data by crosses. Normal interstellar reddening vectors are shown for 
both colors. Clearly, the star does not follow these lines but 
maintains a nearly uniform color while fading by more than 3 mag. At 
minimum light the median color appears to be bluer by $\sim$0.1 mag than its 
out-of-eclipse value.\label{color}}
\end{figure}

During totality, which lasts about 18 days, the system is still visible, 
but at about 5\% of its out-of-eclipse intensity. The spectrum and 
color (Fig. \ref{color}) during this time indicate that we are seeing 
reflected starlight from the circumstellar matter. The knife-edge model 
(Fig. \ref{knifeedge}) includes this component as a constant 
addition at all phases. In fact, the ``central" reversals and short time 
scale fluctuations during minima indicate that variable amounts of reflected 
light are present. More detailed modeling and, in particular, polarization 
measurements should ultimately provide important constraints on the detailed 
distribution of circumstellar matter.  The fact that successive ingress and 
egress data do not match precisely indicates that there are fine scale opacity 
variations in the edge itself. It is also likely that the stellar surface is 
inhomogeneous to some extent, since stars of this mass and age commonly have 
dark (magnetic) or bright (accretion) spots. Detailed photometric and 
spectroscopic monitoring during ingress or egress should reveal much about 
the structural and optical properties of the obscuring matter and perhaps also 
the stellar surface.

Optical photometry in bluer bands (B, V and R) was obtained at a number of 
phases, permitting us to study the color behavior of the star. These data 
indicate, remarkably, that there is no reddening of the star, even as it fades 
by more than 3 magnitudes (see Fig. \ref{color}). This means the eclipsing 
object is, indeed, a very sharp knife-edge with no detectable optically thin 
transition zone or that the optically thin ``atmosphere" is devoid of small 
dust grains (which invariably produce reddening in the interstellar medium). 
During minimum there is large scatter in the color data, owing to the 
faintness of the star, but the median color is bluer by $\sim$0.1 mag than 
when out of eclipse. This presumably indicates that at least some small 
dust grains are present in the circumstellar matter and that we are seeing 
the star partly or entirely by scattered radiation near minimum light. There 
are real variations by up to 20\% in the brightness of the star which occur 
on time scales as short as one hour during minima, indicative of changing 
orientations of the star and scattering clouds. The short time scales imply 
length scales for the variable features of less than 0.01 AU, or one solar 
diameter, assuming they are in Keplerian orbits with periods of 48 or 97 d. 

In an attempt to detect the circumstellar matter of KH 15D by its emission, 
we have obtained near-infrared and millimeter wavelength observations of the 
star at the Infrared Telescope Facility in Hawaii and the Owens Valley Radio 
Observatory millimetre-wave array in California, respectively. The near-infrared 
(JHKL) colors both in and out of eclipse are consistent with an unreddened 
K7 star exhibiting no infrared excess attributable to a disk. The mm wavelength 
observations did not detect a source at the optical position. Unfortunately, 
neither of these results provides a strong constraint on the possible mass of 
a disk. The upper limit on the mm emission translates to a mass limit of about 
7 Earth masses of dust at a temperature of about 500 K, while the lack of 
detectable infrared excess may simply reflect the fact that the putative disk 
is practically edge-on and opaque at these wavelengths.

Spectra with a resolution of $\sim$40000 covering the range 478 - 690 nm 
were obtained on three nights with the Very Large Telescope (VLT) of the 
European Southern Observatory on Mt. Paranal in Chile. Spectra were obtained 
on UT 29 Nov (just prior to ingress), 14 Dec (near central minimum), and 20 Dec 
(during the early part of egress), 2001. A full account of the information 
contained in these data will be given elsewhere \citep{ham02}. They show, however, that the 
spectrum during eclipse is basically an attenuated version of the bright state,
except for the emission lines. Near minimum and during egress, the 
H$\alpha$ equivalent width is substantially larger ($\sim$30 and 50 \AA,
respectively) than when the star is at maximum light ($\sim$2 \AA). 
Essentially, the spectrum makes a transition from one characteristic 
of non-accreting or low accretion rate 
``weak-lined'' T Tauri stars to one characteristic of ``classical''
T Tauri stars (CTTS), 
objects with clear evidence for active accretion of gas from a 
circumstellar disk. In particular, broad wings, extending to several hundred 
km s$^{-1}$ are seen on the H$\alpha$ (and H$\beta$) lines during 
minimum light and during egress, indicating that the 
star may still be in an active accretion stage.  

A cross-correlation of the 29 Nov and 20 Dec spectra yields a radial 
velocity difference (in the sense Nov. minus Dec.)
for the star between those two dates of +3 $\pm$1 km s$^{-1}$. The 
interpretation of this result is complicated by the fact 
that the line profiles change shape during the eclipse, 
presumably due to the selective elimination of direct radiation from 
parts of the star's surface as well as an increased contribution of 
scattered light to the spectrum. \citet{ham02} will explore these issues further. 
Here we note that, if the K7 
star were orbiting the center of mass of the system with a 48.36 d period in 
response to a planet embedded in the obscuring matter, it would have yielded 
a {\it negative} value for the velocity difference on those dates. Therefore, 
assuming that the line profile effects are small, we can 
place a limit of  $\sim$10 Jupiter masses on any mass associated with a 
single obscuring clump orbiting with a 48 d period. With only three spectra 
currently available we cannot, of course, rule out the possibility of 
additional significant mass(es) in the system orbiting with a 97 d (or other) 
period. However, there is no evidence in the spectrum for such a companion so 
it is unlikely to be comparable in mass to the visible K7 star. A radial 
velocity study spanning at least one season with a precision of at least 
1 km s$^{-1}$ is clearly required and is underway in collaboration with 
G. Marcy of U.C. Berkeley. 

\section{Discussion}

KH 15D is a unique and amazing object that promises to tell us much about 
conditions in the inner circumstellar region of a solar-like star of planet-forming age.
The principle purpose of this contribution is to summarize the 
available observations of the star with the hope of stimulating additional 
observations, and to support those by providing an updated ephemeris 
and description of the phenomena. However, a brief discussion of 
possible models is in order, based on the information currently 
available. These fall into two categories, depending on whether the K7 star 
is the dominant mass in the system or not. If it is, then 
the basic model involves occulting matter orbiting the star. If it is 
not (i.e. if KH 15D is a binary star system and the unseen component 
has a mass comparable to or larger than the K7 star), then the observed 
phenomena could be caused wholly or in part by the motion of the visible star 
with respect to a circumbinary disk. We discuss the ``single star'' 
interpretation first, noting that it also applies to binary models
in which the K7 star is the dominant mass.

\begin{figure}
\plotone{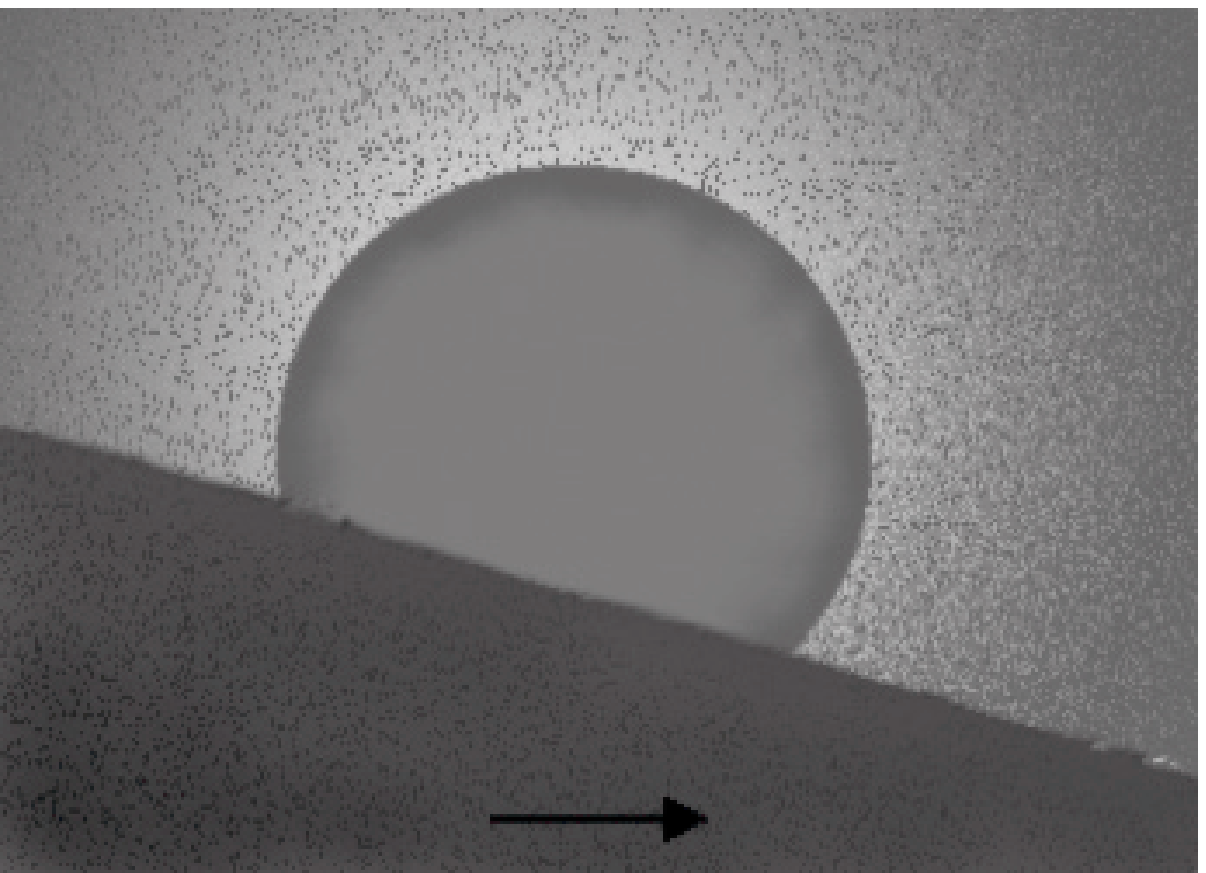}
\caption{A sketch of the knife-edge model during ingress. The top of the 
occulting material is inclined by $\sim$15$\degr$ to its direction of motion, 
which is horizontal in this figure. The star is cooler (and, therefore, redder) 
than the Sun and exhibits limb darkening appropriate to its spectral class. 
Scattered light arises from the visible portion of the circumstellar 
matter and 
(possibly) a second occulting feature on the opposite side of the star. 
The scattered light, which is bluer than direct sunlight, becomes increasingly 
important as the star's photosphere is occulted. \label{model}}
\end{figure}

Assuming that the eclipse is caused by a feature or 
features orbiting the K7 star with a period of 48 or 97 days, 
we can apply Kepler's third law to derive a semi-major axis for the 
orbit of the 
occulting matter of 0.21 or 0.32 AU, respectively. Therefore, if this 
model is correct, we are 
probing a region of the circumstellar environment
closer to the star than Mercury is to the Sun. 
The occulting feature appears to have a very sharp edge, such that 
a knife-edge model fits the data well. However, the transit 
time for an object orbiting at 0.2-0.3 AU from the K7 star is only
$\sim$0.5 d, much less than the time of ingress or egress 
($\sim$2-2.5 d). Therefore, if the star is eclipsed by a sharp-edged 
orbiting feature, its occulting edge
must be inclined by about 15$\degr$ to its direction of motion. 
 A sketch of what this would look like 
during the early stages of ingress is shown in Fig. \ref{model}. Such an 
eclipsing structure could be caused by density waves or a corrugation of 
a 
disk driven by an embedded planet or brown dwarf, as \citet{b00} for example, 
have modeled. Alternatively, the features could be associated with a 
resonance of a yet undetected mass. During the eclipse, the system is seen 
primarily or entirely by reflected 
light from the circumstellar matter. Models such as these have many attractive 
features, including the potential to explain the central reversal as back 
scattering off the wave on the far side of the star or as a local minimum in 
the opacity near the location of an embedded planet or proto-planet. 

If KH 15D is a binary in which the K7 star moves substantially 
with respect to the center of mass of the system, then an entirely different 
explanation of the observed phenomena is possible. Namely, one could 
imagine the K7 star passing above (and, possibly, alternately below) the 
plane of an occulting (presumably circumbinary) disk. In other words, 
it would be primarily the motion of the star, in this model,
that was causing the eclipse, not the 
motion of the occulting matter. As noted 
previously, this appears less likely at present for two reasons. First,
there is only evidence for a single stellar spectrum from the system 
at any phase, even when the photosphere of the K7 star is completely 
occulted. Second, there is very little difference in the radial 
velocity of the K7 stars on two dates separated by 21 d. Neither of these 
arguments is sufficient, however, 
to rule out a binary model. For example, if the visible K7 star were 
on an eccentric orbit about a slightly more massive star and the orbit 
was properly
inclined to the plane of a circumbinary disk, one could reproduce 
most or all of the observations, including the central peaks and the 
ingress/egress time scales. A comprehensive radial velocity 
study, as is under way, is clearly required to make progress. 
We note that, regardless 
of whether KH 15D proves to be a single or binary star, its unique 
orientation provides us with a powerful 
tool for studying the structure and evolution of circumstellar matter close 
to a young star on an unprecedented fine scale.




\acknowledgments

We gratefully acknowledge the support of NASA through its Origins of Solar 
Systems program, Sigma Xi, and Mt. Holyoke College for this work. We also 
thank D. Lin, G. Bryden and M. Kucher for useful conversations regarding 
the interpretation and P. D. Drager for creating Fig. \ref{model}.




\clearpage


\clearpage 

\clearpage 

\clearpage 

\clearpage 

\end{document}